\documentclass[a4paper]{jpconf}
\usepackage{graphicx}
\usepackage{amsmath}
\usepackage{color}
\usepackage{wrapfig}
\newcommand{\refr}{}
\newcommand{\refbbb}{}
\begin{document}
\title{Noise Budget and Interstellar Medium Mitigation Advances in the NANOGrav Pulsar Timing Array}

\author{ 
T.\,Dolch$^{1}$ (for the NANOGrav Collaboration), 
S.\,Chatterjee$^{2}$,
J.\,M.\,Cordes$^{2}$,
P.\,B.\,Demorest$^{3}$,
J.~A.~Ellis$^{4}$, 
M.\,L.\,Jones$^{4,5}$,
M.\,T.\,Lam$^{4,5}$,
T.\,J.\,W.\,Lazio$^{6}$,
L.\,Levin$^{7}$,
M.\,A.\,McLaughlin$^{4,5}$,
N.\,T.\,Palliyaguru$^{8}$,
D.\,R.\,Stinebring$^{9}$
}

\address{$^{1}$Department of Physics, Hillsdale College, 33 E. College Street, Hillsdale, MI 49242, USA}
\address{$^{2}$Astronomy Department, Cornell University, Ithaca, NY 14853, USA}
\address{$^{3}$National Radio Astronomy Observatory, 1003 Lopezville Rd., Socorro, NM 87801, USA}
\address{$^{4}$Department of Physics and Astronomy, West Virginia Univ., Morgantown, WV 26506, USA}
\address{$^{5}$Center for Gravitational Waves and Cosmology, West Virginia University, Chestnut Ridge Research Building, Morgantown, WV 26505}
\address{$^{6}$Jet Propulsion Laboratory, California Institute of Technology, 4800 Oak Grove Drive, Pasadena, CA 91106, USA}
\address{$^{7}$Jodrell Bank Centre for Astrophysics, School of Physics and Astronomy, The University of Manchester, Manchester M13 9PL, UK}
\address{$^{8}$Department of Physics, Texas Tech University, Box 41051, Lubbock, TX 79409, USA}
\address{$^{9}$Department of Physics and Astronomy, Oberlin College, Oberlin, OH 44074, USA}
\ead{tdolch@hillsdale.edu}

\begin{abstract}
Gravitational wave (GW) detection with pulsar timing arrays (PTAs) requires accurate noise characterization. The noise of our Galactic-scale GW detector has been systematically evaluated by the Noise Budget and Interstellar Medium Mitigation working groups within the North American Nanohertz Observatory for Gravitational Waves (NANOGrav) collaboration. Intrinsically, individual radio millisecond pulsars (MSPs) used by NANOGrav can have some degree of achromatic red spin noise, as well as white noise due to pulse phase jitter. Along any given line-of-sight, the ionized interstellar medium contributes chromatic noise through dispersion measure (DM) variations, interstellar scintillation, and scattering. These effects contain both red and white components. {\refr In the future, with wideband receivers, the effects of frequency-dependent DM will become important.} Having anticipated and measured these diverse sources of detector noise, the NANOGrav PTA remains well-poised to detect low-frequency GWs.
\end{abstract}

\vspace*{-2em}
\section{Introduction}

The NANOGrav PTA collaboration currently times 71 MSPs (as of 2018 January 1) in a long-term campaign with the Arecibo Observatory (AO) and the Green Bank Telescope (GBT). {\refr Each pulsar is timed roughly monthly, with a small number of high-precision pulsar timed weekly}, by one or both telescopes for $\sim$20\,min per observation. The pulse profiles (intensities as a function of pulse phase, radio frequency, and time during the observation) are ``folded'', or time-averaged over {\refr a} number of pulse periods, resulting in high S/N times-of-arrival (TOAs) when compared with a local high-precision hydrogen maser time standard. {Within NANOGrav, the various working groups function together to maintain long-term pulsar timing and produce its data products. Outside NANOGrav, the collaboration's efforts are joined with the European Pulsar Timing Array (EPTA) collaboration and the Parkes Pulsar Timing Array (PPTA) collaboration as the International Pulsar Timing Array (IPTA).}

As a National Science Foundation Physics Frontiers Center, the NANOGrav collaboration produces periodic public data releases. {\refr Each release (including associated metadata) is based on the deterministic timing model that can be updated as a result of the NANOGrav observations. The residuals between the observed and predicted TOAs} (utilizing the uncertainties in timing residuals, also known as ``RMS residuals'') would become significantly non-zero in the presence of unmodeled physical processes. As for the signature of GWs in pulsar timing residuals, their significance in an individual pulsar may be low. However, {\refr GWs with light-year wavelengths passing through} solar system should perturb TOAs from pulsars, each with a different line-of-sight (LoS), in a correlated manner \cite{hd83}. The residuals from the NANOGrav dataset, begun in 2005, are sensitive to GWs in the nHz regime of the GW spectrum. The most likely GW source expected for PTA detection is a stochastic background of GWs from merging supermassive black hole binaries (SMBHBs), although continuous wave (CW) sources from individual SMBHBs in pre-merger orbits are also a possibility, as are memory bursts from the instants of SMBHB merger. {\refr In addition to SMBHBs, cosmic strings are also a potential GW source.}

Each pair of pulsars in the PTA, in principle up to {$71 \choose 2$} pairs, is analagous to the two arms of a laser interferometer GW detector. Just as a ground-based detector has its own noise spectrum as a function of GW frequency, so does every pulsar have its noise spectrum. Because a PTA is a Galaxy-sized GW detector, noise and signal must be measured together. Each NANOGrav data release now includes updated noise models, {\refr jointly derived by} fitting all the pulsars' data from 2005 to the present. Some parameters correspond to individual pulsars, and others correspond to the Earth's observing conditions (for example, the solar system ephemeris model, which determines the barycentric position of the observatory when a particular observation occurs). Fitting for all-PTA noise models has been a primary activity of the Detection and Timing working groups within NANOGrav.

The Noise Budget (NB) working group provides an additional approach to noise characterization, in which each pulsar's noise spectrum is understood as a sum of contributions from various physical process appropriate to that pulsar. Just as all-PTA noise fits create a noise spectrum from the 71 pulsars jointly, individual noise measurements characterize a noise spectrum for a particular pulsar. This second measurement method approaches the ideal of an in-situ measurement of the noise in each detector component. In particular, noise processes can be characterized as either chromatic or achromatic. Growing achromatic noise in multiple pulsars, for example, would be consistent with a stochastic GW background. Chromatic noise, however, is consistent with unidentified processes affecting pulsar signal propagation in the ionized interstellar medium (IISM), which are radio-frequency dependent, and therefore more easily mitigated. Additional, the NB group can identify long-term observing strategies that can minimize noise. 

Measuring and ultimately mitigating noise due {\refr to} the IISM is under the purview of NANOGrav's Interstellar Medium Mitigation (IMM) group. The primary way that the IISM affects radio propagation is through dispersive delays, in which for a particular pulsar, TOA $\propto 1/\nu^2$, where $\nu$ is the radio frequency of received pulsar emission. The delay is proportional to DM, which is a measurement of the number of electrons along the LoS. The effects of dispersion are removed from the data in real time with the coherent dedispersion algorithm in pulsar backends; however, this removal assumes a particular DM value. DM for each pulsar varies {\refr by one part in 1000} of a fiducial DM in a timing model, enough to affect TOAs on timescales of $\mu$s or greater. NANOGrav observations are carried out at 2 or 3 center frequencies per pulsar, so that DM {\refr can be measured at every epoch}. In addition to DM variations, diffractive interstellar scintillation (DISS) through a single thin {\refr phase-screen} IISM overdensity perturbs TOAs with a $1/\nu^{4.4}$ dependence for an IISM distribution obeying a Kolmogorov turbulence distribution. 
 
\begin{figure}[h]
\begin{minipage}{18pc}
\vspace*{-3.5em}
\includegraphics[width=18pc]{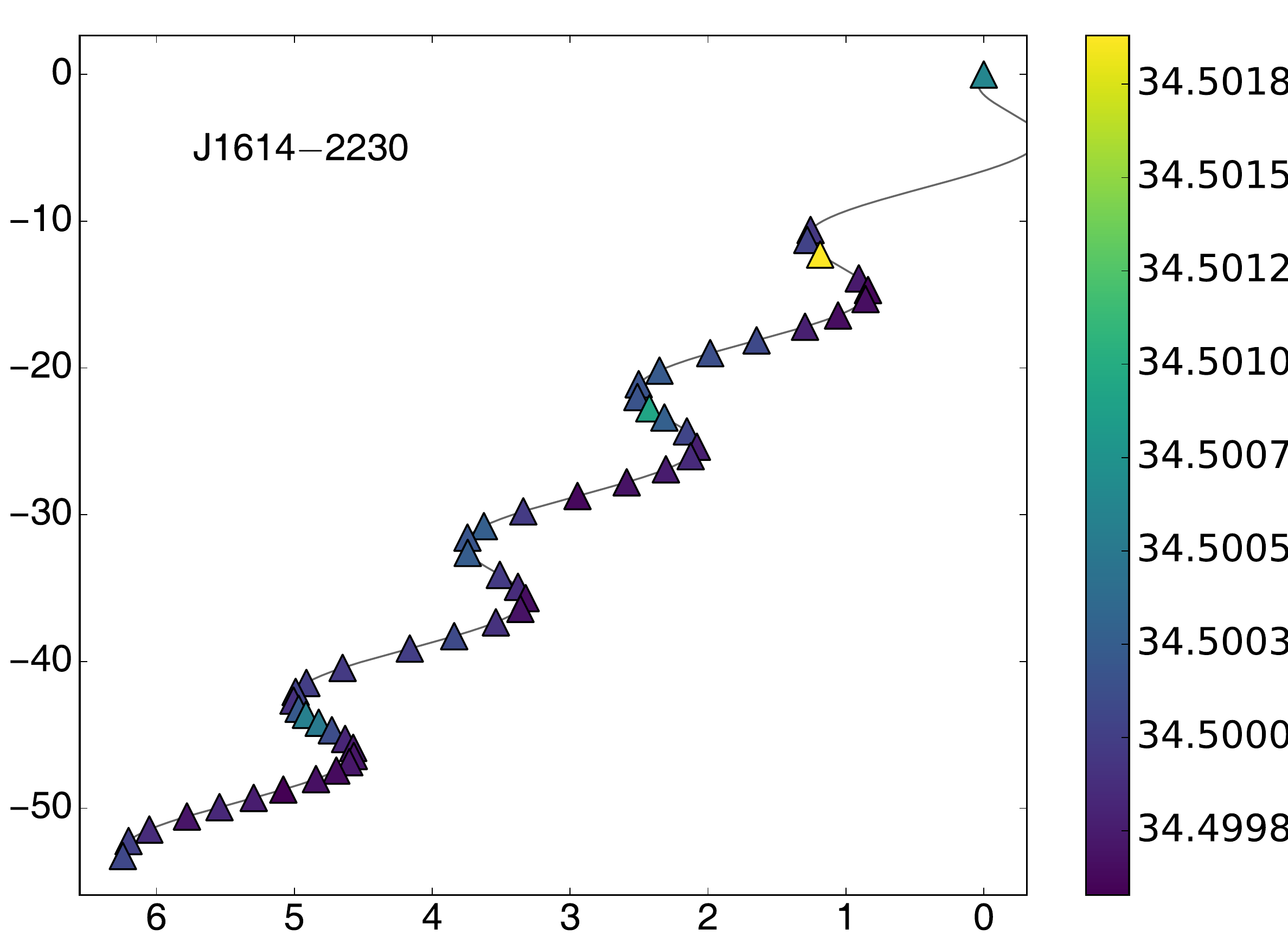}

\caption{{\refr Example plot from \cite{jones17}, showing the right ascension and declination of PSR J1614--2230, beginning at the origin. Within the plotted positions, the color shows the measured DM, giving a picture of the IISM structure. Annual periodicities in DM are visible, probably due to a particular small-scale overdensity.}}

\end{minipage}
\begin{minipage}{18pc}
\vspace{-5em} 
\includegraphics[width=18pc]{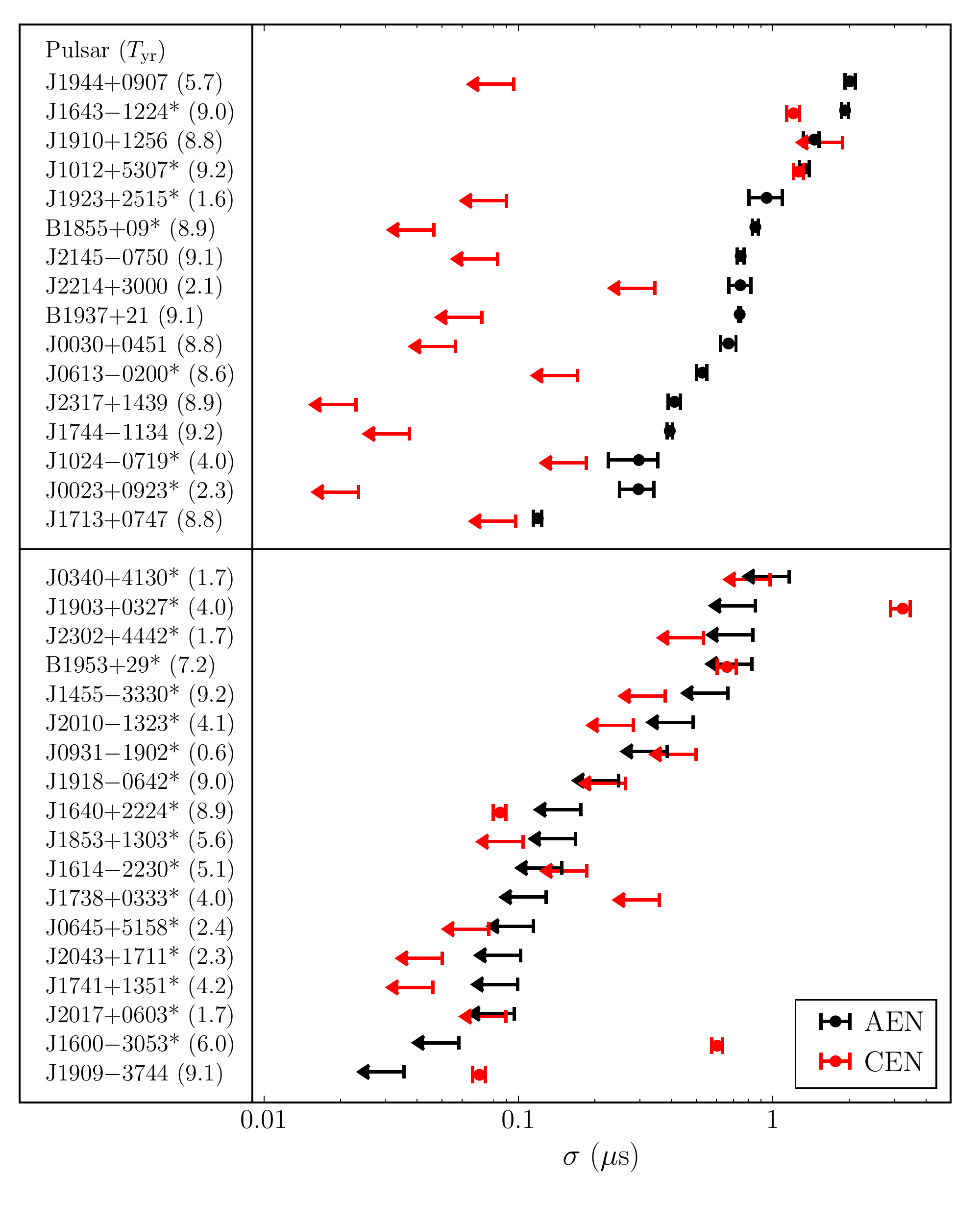}
\vspace*{-3em}
\caption{{Timing residual noise measured in \cite{lam17a} for all 9-year NANOGrav pulsars. Achromatic noise is in black and chromatic noise is in red. Pulses are sorted from the top according to their achromatic noise values. The top panel shows pulsars with achromatic noise detections, and pulsars with achromatic noise limits are on the bottom panel. The left column shows the number of years a pulsar has been observed by NANOGrav in the 9-yr dataset. Upper limits are shown by asterisks.}}
\end{minipage} 
\end{figure} 
 
\vspace*{-1em} 
 
\section{Advances in Interstellar Medium Mitigation}
 
The IMM working group has analyzed potential effects of the IISM that would not ordinarily be characterized in the standard process of TOA and timing residual generation. Timing models are inherently deterministic. Variations in DM are not generally predictable, and it is a standard PTA practice to measure the DM every epoch utilizing TOAs at widely-spaced frequencies. DM variations do, however, have both deterministic and stochastic components, the predictable component arising from the deterministic LoS change due to the Earth's and a pulsar's velocities; see the IMM group's results in \cite{jones17} (Figure 1; which is the example of PSR J1614--2230) and \cite{lam16a}. Multipath diffractive scattering also occurs in addition to DM delays, when an overdensity in the IISM temporarily redirects the signal's propagation paths. The diffraction creates {\refr a changing interference pattern} across observing bandwidth, and also {\refr forward scatters a pulse's flux in time with scattering timescale $\tau_{\mathrm{d}}$}. Scattering timescales are nontrivial to measure. {\refr A constant $\tau_{\mathrm{d}}$ would be irrelevant for GW detection.} If significantly varying, however, the scattering tails of received pulses would provide another source of stochastic noise in timing residuals \cite{stine13}. In principle, re-measuring a pulsar's scattering tail {\refr at} every epoch is possible, but a current implementation is not computationally trivial across the entire backend receiver bandwidth (at AO, the Puerto Rican Ultimate Pulsar Processing Instrument, PUPPI; at the GBT, the Green Bank Ultimate Pulsar Processing Instrument, GUPPI; \cite{duplain}). 

In the NANOGrav nine-year data release \cite{a15b}, the (then) 37 pulsars were shown to generally not have significant variations in scattering parameters \cite{levin}, compared to the RMS residuals from all noise processes. Scattering timescales at each observing epoch were computed not from directly measuring scattering tails, but by utilizing dynamic spectra (DS; signal intensity as a function of both time and radio frequency). The interference pattern across bandwidth due to the multipath scattering of the on-pulse signal at any given time is directly related to $\tau_{\mathrm{d}}$ according to $\tau_\mathrm{d} \sim 1/(2\pi{\Delta{\nu_\mathrm{d}}})$, where $\Delta{\nu_\mathrm{d}}$ is the diffractive bandwidth. The constructive interference peaks (or ``scintles'' with a characteristic width equal to $\Delta{\nu_\mathrm{d}}$) change on diffractive timescales. The unpredictable filling pattern of scintles in a DS is a source of white noise {\refr in timing residuals, $\sigma_{\mathrm{DISS}}$}. Scattering timescale characterization using dynamic spectra (as opposed to IISM characterization through TOAs alone) is a good example of using ancillary data in PTA science with the aim of GW detection (see also \cite{keith13}).

The low level of scattering variation was to some degree a selection effect, because a pulsar with highly varying $\tau_{\mathrm{d}}$ values would not have originally been selected for inclusion in a PTA. {\refr In future} data releases, however, sensitivity to GWs will have increased, such that the $\tau_{\mathrm{d}}$ measurements in \cite{levin} may need to be corrected through techniques such as cyclic spectroscopy ({\refr \cite{pal15}, \cite{dem11}, \cite{w13}, \cite{arch14}}), which is another project within the IMM group. 
 
Another possible unquantified stochastic noise process could arise from more complex scattering processes detectable with very wide-bandwidth receivers. The analysis in \cite{css} shows that highly-scattered pulsar signals may, along different ray paths, propagate through significantly different numbers of electrons, integrated along their respective paths, before their refocusing. As a result, over very wide bandwidths, pulsar signals might follow a different law than TOA $\propto 1/\nu^2$ (because otherwise, the standard cold-plasma propagation law assumes an essentially 1D propagation). This prediction is not only significant for future wideband receivers, but also an interesting explanation for the excess chromatic noise found in many NANOGrav pulsars (to be described in the following section). In the future, a more complex frequency-dependent propagation law might be a standard component of timing models.  

\section{Advances in Noise Budget Descriptions}

The NB working group has constructed bottom-up noise models for each individual pulsar. It is essential to have such an approach along with top-down, all-PTA global noise modeling; a stochastic background of GWs would initially show up as a source of red noise in the PTA as a whole. However, individual-pulsar noise modeling, advancing significantly with each data release, becomes increasingly useful as both a prior on all-PTA noise modeling, as well as an independent pipeline to characterize the accumulation of red noise in pulsars.

Red noise can be characterized \cite{sc2010} as having chromatic and achromatic components. Stochastic noise processes intrinsic to a pulsar such as spin noise, intrinsic to the observation position such as noise from ephemeris model uncertainties, or even extrinsic ``red noise'' from a GW background (more properly, a signal with a red power spectrum) have no radio frequency dependence. Stochastic noise processes arising from the pulsar signal's interaction with the IISM should have a radio frequency dependence. {\refr In a} recent study from the NB group \cite{lam17a}, {\refr we conducted} measurements of red noise in each NANOGrav pulsar, separating the noise out into chromatic and achromatic components.

The red noise measurements in \cite{lam17a} (Figure 2) are also referred to as ``excess noise'', in other words, noise in excess of well-characterized white noise processes. The white noise models come from \cite{lam16b} (Figure 3), in which bottom-up white noise measurements are conducted on NANOGrav pulsars. The global PTA fit yields three white noise components to TOAs from the \textsc{Tempo2} software package: ``EFAC'', which simply multiplies each TOA uncertainty by a scale factor; ``EQUAD'', which adds a white noise component in quadrature to each TOA uncertainty (and this performed independently at each receiver frequency) and ``ECORR'', which adds the same white noise component to each receiver frequency, in other words, as achromatic noise.

In the top-down approach, significant ECORR values can be likely identified with a well-known source of achromatic white noise in individual pulsars: pulse phase jitter. However, the white noise modeling in \cite{lam16b} can use the residuals from all epochs to directly measure the white noise due to pulse phase jitter in each pulsar, by fitting for the white noise in the RMS residual values that does not vanish even in the highest S/N residuals. This method was first applied in \cite{d14} and then extended to all NANOGrav pulsars in \cite{lam17b}. Additionally, \cite{lam16b} utilizes the scattering measurements in \cite{levin} (deriving from DS not utilized in a global PTA fit) to characterize $\sigma_{\mathrm{DISS}}$ for each pulsar, separating $\sigma_{\mathrm{DISS}}$ out from pulse phase jitter explicitly.

Having characterized the white noise for each pulsar individually, \cite{lam17a} then computes the excess red-noise not described by the white noise models (Figure 3). This results in either noise upper limits or significant measurements depending on the pulsar. {\refr We find that} 16 of the 37 nine-year data release pulsars are found to have achromatic excess noise at the level of 100\,ns or greater. {\refr We also find that 7 pulsars} have significant chromatic excess noise. Taking the chromatic measurements and upper limits together, and plotting each pulsar's noise value (or limit) with respect to its DM, an approximate proportional relationship emerges, albeit with outliers. The unmodeled chromatic noise may be due to either $\tau_{\mathrm{d}}$ variation (as in \cite{levin}) and/or due to additional radio frequency-dependencies such as that proposed in \cite{css}.

\begin{wrapfigure}{c}{0.6\textwidth}
\begin{center}
\includegraphics[width=0.6\textwidth]{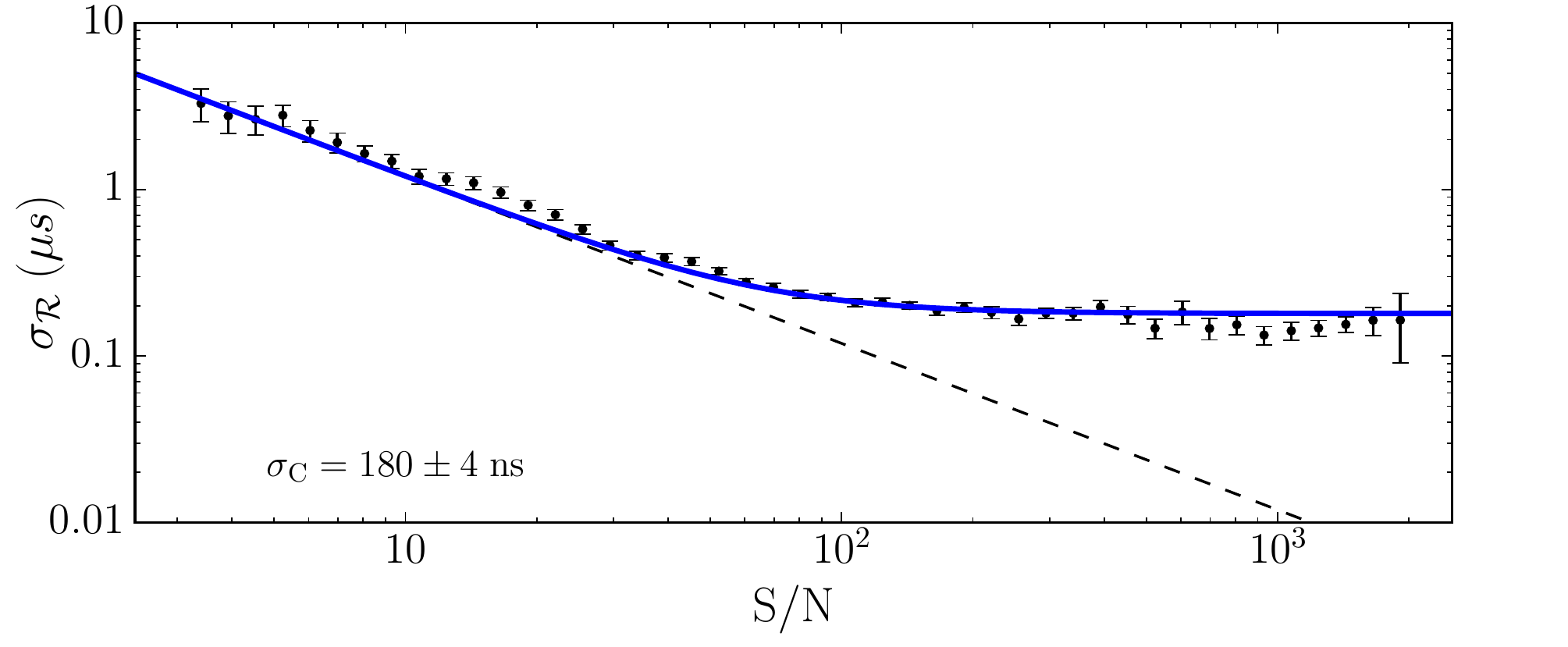}

\caption{{\refr PSR J1713+0747's timing residuals, from \cite{lam16b}. Data are from Arecibo Observatory in the NANOGrav 9-year data release at a center frequency of 1400\,MHz. The white noise model used for the 9-year NANOGrav dataset predicts an RMS-residual vs. residual S/N shown by the dashed line. The solid line uses the data to fit for an additional term in pulse phase jitter, which is measured for NANOGrav 9-year pulsars in \cite{lam16b}.}}


\end{center}
\end{wrapfigure}

As for the excess achromatic noise, there are a number of plausible noise mechanisms: low-level, unmitigated, broadband radio frequency interference (RFI); polarization calibration errors; and yet-umodeled wandering of planetary ephemeris parameters. An intriguing possible explanation for the growing red noise measurements in the nine-year data release (as compared to the five year data release \cite{dem13}) is the presence of a GW background. However, a convincing eventual detection would also include \emph{correlated} noise \cite{hd83} between pulsars in addition to a common red signal. With additional NANOGrav data releases utilizing both AO and the GBT, the further dissection of achromatic noise into the possible components described here is highly likely.

In order to best make {\refr use} of future NANOGrav observing time, another project of the NB working group \cite{lam17b} has been the determination of the optimal observing center frequencies and bandwidths for each pulsar, for each telescope. Currently, the standard timing data from the GUPPI and PUPPI backends is taken at up to {\refr three} widely separated center frequencies. Future receivers are expected to span an octave of bandwidth or more. The study in \cite{lam17b} applies to a future wide-bandwidth receiver, suggesting its optimization for a specific PTA, given the varieties of frequency-dependent IISM noise in timing residuals present across the most significant NANOGrav pulsars.

{\refr Studying the NB of MSPs also results in ancillary science. The 24-hr global campaign \cite{d14} was a 9-telescope observation by the IPTA conducted for the purpose of studying PSR~J1713+0747's noise budget on multiple-hour timescales. This particular MSP was especially important for noise characterization as the most sensitive individual pulsar observed by the entire IPTA. As a result of the campaign, a new GW limit was obtained \cite{d16} in the GW frequency range 10$^{-5}$ -- 10$^{-3}$\,Hz, both for the entire sky and for any GW sources in the direction of the pulsar.}

\section{Conclusions}

The efforts of the NB and IMM working groups have characterized the noise, largely due to IISM effects, of the pulsars in the NANOGrav PTA as a whole. These results include models of the IISM and its interaction with pulsar signals (\cite{lam16a}, \cite{jones17}). In particular, the models of \cite{jones17} are applicable to future datasets, for which the deterministic and stochastic components of DM variations can be more confidently separated. Direct white noise measurements of \cite{lam16b} and \cite{levin} enable a high-confidence identification of intraday noise sources with specific physical processes, leading in turn to a measurement of each pulsar's excess noise, which can be further decomposed into chromatic red noise and achromatic red noise. The achromatic red noise component may itself already contain timing perturbations due to a stochastic GW background, as NANOGrav approaches the sensitivity in which predicted background levels should be present \cite{sesana17}. {\refr However, more definite statements will require discriminating GW signatures from those of other red noise processes}, highlighting the importance of continual noise budgeting. The value of continuing NB and IMM characterization is not only for GW detection; indeed, the ancillary ISM science can reveal unusual events present in DM(t) data, such as that reported by the PPTA in PSR~J1713+0747 \cite{keith13}. Such events inform plasma lensing models which are also relevant to studying propagation effects in fast radio bursts \cite{cordes17}.

\section{Acknowledgments}

{\refr The NANOGrav project receives support from National Science Foundation (NSF) Physics Frontier Center award number 1430284 and NSF-PIRE program award number 0968296.} TD acknowledges travel support from the NANOGrav PFC award. Part of this research was carried out at the Jet Propulsion Laboratory, California Institute of Technology, under a contract with NASA. JAE acknowledges support by NASA through Einstein Fellowship grant PF4-150120. {\refbbb We thank the anonymous referee.} {\refr This work made use of NASAÕs Astrophysics Data System Abstract Service and the arXiv.org astro-ph preprint service.}

\smallskip

\end{document}